\documentstyle[epsf]{mn}
\newcounter{saveeqna}

\newcounter{saveeqnb}
\newcommand{\alpheqntext}
{\setcounter{saveeqnb}{\value{equation}}%
\stepcounter{saveeqnb}%
\setcounter{equation}{0}%
\renewcommand{\theequation}
{\mbox{\arabic{saveeqnb}.\alph{equation}}}}
\newcommand{\reseteqntext}
{\setcounter{equation}{\value{saveeqnb}}%
\renewcommand{\theequation}{\arabic{equation}}}
\newcommand{\sig}{\:\lower0.6ex\hbox{$\stackrel{\textstyle >}{\sim}$}\:}
\newcommand{\sil}{\:\lower0.6ex\hbox{$\stackrel{\textstyle <}{\sim}$}\:}
\newcommand{\sigs}{\:\lower0.4ex\hbox{$\stackrel{\scriptstyle
      >}{\scriptstyle \sim}$}\,}
\newcommand{\sils}{\:\lower0.4ex\hbox{$\stackrel{\scriptstyle
      <}{\scriptstyle \sim}$}\,}
\begin{document}
%***************************************************************

\title{GRAPESPH with Fully Periodic Boundary Conditions:
Fragmentation of Molecular Clouds}

\author[R.~Klessen]{Ralf Klessen\\
Max-Planck-Institut f{\"u}r Astronomie, K{\"o}nigstuhl 17, 69117
Heidelberg, Germany}

%%%%%%%%%%%%%%%%%%%%%%%%%%%%%%%%%%%%%%%%%%%%%%%%%%%%%%%%%%%%%%%%%%%%%%
\maketitle
\begin{abstract}
A method of adapting smoothed particle hydrodynamics ({\sc SPH})
with periodic boundary conditions for use with the special purpose
device {\sc Grape}  is presented.   {\sc Grape} (GRAvity PipE) solves the Poisson and
force equations for an N-body system by direct summation on a specially
designed chip and in addition returns the neighbour list for each
particle. Due to its design, {\sc Grape}  cannot treat periodic particle distributions
directly. This limitation of {\sc GrapeSPH}  can be overcome by
computing a correction 
force for each particle due to periodicity (Ewald correction) on the
host computer using a PM-like method.

This scheme is applied to study the fragmentation process in giant
molecular clouds. Assuming a pure isothermal model, we follow the
dynamical evolution in the interior of a
molecular cloud starting from an Gaussian initial density
distribution to the formation of selfgravitating clumps until most of the gas
is consumed in these dense cores. Despite its
simplicity, this model can
reproduce some fundamental   properties of observed molecular clouds, like  
a clump mass distribution of the form $dN/dm \propto m^n$, with
$n\simeq -1.5$.

%These reveal a hierarchy of clumps and structure on all
%scales accessible by todays telescopes. 
\end{abstract}
%%%%%%%%%%%%%%%%%%%%%%%%%%%%%%%%%%%%%%%%%%%%%%%%%%%%%%%%%%%%%%%%%%%%%%

%%%%%%%%%%%%%%%%%%%%%%%%%%%%%%%%%%%%%%%%%%%%%%%%%%%%%%%%%%%%%%%%%%%%%%
%%%%%%%%%%%%%%%%%%%%%%%%%%%%%%%%%%%%%%%%%%%%%%%%%%%%%%%%%%%%%%%%%%%%%%

\section{Introduction}
In the early 1990's, the first {\sc Grape} boards became available to
the scientific community. {\sc Grape} is an acronym for ``GRAvity
PipE'' and is a hardware device to solve the Poisson and force
equations for an N-body-system by direct summation on a specially
designed chip, thus leading to considerable speed-up in computing this
type of problems (Sugimoto et al. 1990, Ebisuzaki et
al. 1993). Besides the gravitational forces, {\sc Grape} also returns
the list of neighbours for each particle, which makes it attractive
for use in combination with smoothed particle hydrodynamics, {\sc
SPH}, (for a general review on SPH see Benz 1990 or Monaghan 1992, for
combination with the hierarchical tree method, see Hernquist \& Katz
1989, and for implementing with {\sc Grape}, see Umemura et al. 1993,
and Steinmetz 1996). In the last five years, {\sc Grape} boards have
been used for studying a variety of astrophysical problems, ranging
from following the evolution of binary black holes (Makino et
al. 1993a) up to the formation of galaxy clusters in a cosmological
context (Huss, Jain \& Steinmetz 1997). A complete overview is given
in the review article by Makino \& Hut (1997).

Due to its restricted force law, {\sc Grape} cannot treat periodic
particle distributions. This limits its applicability to strongly self
gravitating systems. However, many astrophysical problems require a
correct treatment of periodic boundary conditions. Simulating the
dynamical evolution and fragmentation in the interior of a molecular
cloud, a process that finally leads to the formation of new stars,
requires periodic boundaries to prevent the whole object from
collapsing to the center. The situation is quite similar in
cosmological large scale structure simulations, where periodic
boundaries mimic the homogeneity and isotropy of the initial matter
distribution.  In the following, we present a method to overcome this
limitation and to incorporate fully periodic boundary conditions into
an N-body or SPH algorithm using {\sc Grape}.  The basic idea is to
compute a periodic correction force for each particle on the host
computer, applying a particle-mesh (PM) like scheme: We first compute
the forces in the isolated system using direct summation on {\sc
Grape}, then we assign the particle distribution to a mesh and compute
the correction force for each grid point, by convolution with the
adequate Green's function in Fourier space. Finally we add this
correction to each particle in the simulation. The corrective Green's
function can be constructed as the offset between the periodic
solution (calculated via the Ewald (1921) approximation) and the
isolated solution on the grid.

Our approach is thus related to the method described in Huss
et al. (1997). However,  they apply the combined force calculation
via a periodic PM scheme and direct summation on {\sc Grape} only to
the particle distribution in the center of their simulation
volume. In this region, they compute the 
dynamical evolution of a galaxy cluster with high resolution, whereas
the remaining part of the volume is treated by a pure PM scheme and
provides the tidal 
torques for the cluster in the center. For this reason, they do not have 
to handle possible force misalignment in the border region of the simulation
cube. In our situation, studying the evolution and fragmentation in the
interior of molecular clouds, we need high resolution and periodicity
throughout the {\em whole} simulation volume. Ensuring numerical
stability throughout the complete volume requires additional
care in combining both methods, the PM scheme and {\sc Grape}.

This paper is organised as follows: The next section gives a brief
overview of the main properties of the special purpose hardware {\sc
Grape}; then follows a description of the Ewald method which leads to
the construction of a Green's function for a periodic correction. This
Green's function is applied in a PM like scheme, to obtain a
(corrective) force term which is added to the particle forces in {\sc
GrapeSPH}. Section \ref{opt-test} investigates the performance of the
method in selected test cases and suggests ways of optimising the
algorithm. The next section applies the above to a preliminary
study of the  fragmentation process in the interior of molecular
clouds and the section \ref{summary} finally concludes with a summary
of this paper.

\section{{GRAPE} Specifications}
\label{grape}
{\sc Grape} is a special purpose hardware device, which calculates
the forces and the potential in the gravitational N-body problem by
direct summation on a specifically designed chip with high efficiency 
(Sugimoto et al. 1990, Ebisuzaki et al. 1993). 
We use the currently distributed version, {\sc
Grape-3AF}, which contains 8 chips on one board and therefore can
compute the forces on 8 particles in parallel.  The board is connected
via a standard VME interface to the host computer, in our case a
SUN workstation. C and FORTRAN libraries provide the software
interface between the user's program and the board. The
computational speed of {\sc Grape-3AF} is approximately $5\,$Gflops.

The force law is hardwired to be a Plummer law,
\begin{equation}
{\bf F}_i \:=\: - G \sum_{j=1}^N \frac{m_im_j({\bf r}_i-{\bf r}_j)}{(|{\bf
r}_i-{\bf r}_j|^2 + \epsilon_i^2)^{3/2}}\:.
\end{equation}
Here $i$ is the index of the particle for which the force is calculated
and $j$ enumerates the particles which exert the force; $\epsilon_i$ is
the gravitational smoothing length of particle $i$;  $G$ and $m_i$ and 
$m_j$ are Newton's constant and the particle masses, respectively.
To perform the force calculation for a single particle,
 {\sc Grape-3AF} needs $N+19$ clock
cycles. To achieve a high speed, concessions in the accuracy of the
force calculations had to be made: {\sc Grape} internally works with a
20 bit fixed point number format for particle positions, with a 56 bit
fixed point number format for the forces and a 14 bit logarithmic
number format for the masses (Okumura et al. 1993). Conversion to and
from that internal number representation is handled by the interface
software. The number format limits the spacial resolution in
a simulation and constrains the force accuracy. However, for
collisionless N-body systems, the forces on a single particle
typically need not be known 
better than up to an error of about one percent. In that respect, {\sc
Grape} is comparable to the widely used {\sc Treecode} schemes (e.g. Barnes
\& Hut 1986). Besides the low precision {\sc Grape} board, there is a
high precision version available, {\sc Harp},
% (Hermite AcceleratoR Pipeline), 
which was specifically designed for treatment of collision
dominated systems (Makino, Kokubo and Taiji 1993).

In addition to integrate the time evolution of an N-body system by
direct summation, {\sc Grape} has been incorporated into a {\sc
Treecode} algorithm (Makino \& Funato 1993) and, more recently, into a
P3M code (Brieu, Summers \& Ostriker 1995). Besides forces and
potential, {\sc Grape} returns the list of neighbours for each
particle $i$ within a specified radius $h_i$. This feature makes it
attractive for use in smoothed particle hydrodynamics (Unemura 1993),
where gas properties are derived by averaging over a localised
ensemble of discrete particles. For particles near the surface of the
integration volume, ``ghost'' particles are created to correctly
extend the neighbour search beyond the cell border; no forces are
computed for these particles. For periodic boundaries, the ghosts are
replicas of particles from the opposite side of the cell. Following
the strategy described in Steinmetz (1996), we have modified our SPH
code (Benz 1990, Bate et al. 1995) to incorporate {\sc Grape} and
furthermore added the ability to handle periodic boundary conditions.

Note, that there also has been an attempt to treat periodic boundaries
on a special purpose hardware device, {\sc Wine}, using the Ewald
method hardwired on a special chip (Fukushige et al. 1993).  This work
has been successful. Fukushige et al. (1996) report the development of
the {\sc MD-Grape}, which can handle an arbitrary central force law,
including the Ewald method.

\section{Implementing Periodic Boundaries into {GRAPE}}
The general concept of incorporating periodic boundaries into {\sc
Grape} is the following: We use the Ewald (1921) method to optain a
Green's function that accounts for the correction to potential and
force for a system transformed from having vacuum boundary conditions
to periodic ones. {\sc Grape} computes the forces and potential for
the isolated system. The Green's function is then used in a PM like
scheme to assign a correction term due to periodicity to each particle
in the simulation.

\subsection{Periodic Force Correction -- The Ewald Method}
In 1921, P. Ewald suggested a method to compute the forces in a
periodic particle distribution (he aimed explicitely to compute the
potential in atomic lattices in solids). For a more recent discussion
see Hernquist, Bouchet \& Suto (1991); for applications to large scale
structure simulations see Katz, Weinberg \& Hernquist (1996), or
Dav{\'e}, Dubinski \& Hernquist (1997).

Poisson's equation for a system of $N$ particles which are
infinitely replicated in all directions with period $L$ is (${\bf n}$
being an integer vector)
\begin{equation}
\nabla^2\phi({\bf r}) \:=\: 4\pi G \rho({\bf r})\:=\: 4 \pi G \sum_{\bf n}
\sum_{j=1}^N m_j \delta({\bf r} -{\bf r}_j - {\bf n}L)
\end{equation} 
and can be solved in general using the appropriate Green's function
$\cal G$:
\alpheqntext
\begin{eqnarray}
\label{green}
\phi({\bf r}) & \!\!\! = \!\!\! & -G \sum_{\bf n} \sum_{j=1}^N m_j
{\cal G}({\bf r} -{\bf r}_j - 
{\bf n}L)\\
 & \!\!\! =\!\!\! &  -\frac{G}{L^3}\sum_{j=1}^N\sum_{\bf k} \hat{\cal G}({\bf
k})m_j\exp\left [i{\bf k} ({\bf r} - {\bf r}_j) \right ],
\end{eqnarray}
\reseteqntext
which is for the gravitational
potential  ${\cal G}({\bf r}) = 1/r$, 
 or $\hat{\cal G}({\bf k}) = 4\pi / {k^2}$ in Fourier space. The sum
in Eqn.~3 converges very slowly, which strongly limits its
numerical applicability. 
However, Ewald (1921) realised, that convergence can be improved
considerably by splitting the 
Green's function into a short range ${\cal G}_S$ and a long range part
${\cal G}_L$ and solving
the first in real space and the latter one in Fourier space:
\alpheqntext
\begin{eqnarray}
{\cal G}_S({\bf r}) = \frac{1}{r}{\rm erfc}(\alpha r) &
\!\!\!\!;\!\!\!\! &
\hat{\cal G}_S({\bf k}) = \frac{4\pi}{k^2}\left[
1-\exp\left(-\frac{k^2}{4\alpha^2}\right)\right]\\
{\cal G}_L({\bf r}) = \frac{1}{r}{\rm erf}(\alpha r) &
\!\!\!\!;\!\!\!\! & \hat{\cal  
G}_L({\bf k}) =
\frac{4\pi}{k^2}\exp\left(-\frac{k^2}{4\alpha^2}\right)\:. 
\end{eqnarray}
\reseteqntext
Here $\alpha$ is a scaling factor in units of inverse length and 
${\rm erf}(x)$ is the error function with ${\rm
erfc}(x)$ its complement,
\begin{equation}
{\rm erf}(x) \:=\: \frac{2}{\sqrt{\pi}} \int_0^x e^{-\xi^2}d \xi \;.
\end{equation}
With the Green's function split into two parts, ${\cal G} = {\cal G}_S
+ {\cal G}_L$, the potential reads as 
$\phi({\bf r}) =\phi_S({\bf r}) + \phi_L({\bf r})$:
\begin{eqnarray}
\label{potential}
\phi({\bf r}) &\!\!\!=\!\!\!& -G \sum_{j=1}^N m_j \left[ \sum_{\bf n} \frac{{\rm
erfc}(\alpha|{\bf r} - {\bf r}_j - {\bf n}L|)}{|{\bf r} - {\bf r}_j -
{\bf n}L|} \right. - \nonumber \\
 & & - \left. \frac{1}{L^3}\sum_{\bf k} \frac{4\pi}{k^2} \exp
\left( -\frac{k^2}{4\alpha^2} \right) \cos\big( {\bf k} ({\bf r} - {\bf
r}_j) {}_{}\big) \right]. 
\end{eqnarray}
%\newpage\noindent 
Finally, the force (say exerted onto particle $i$) is
\begin{equation}
{\bf F}_i \:=\: {\bf F}({\bf r}_i) \:=\: - m_i \nabla \phi({\bf r}_i) \:=\: G m_i
\sum_{j \not= i}m_j {\bf f}({\bf r}_i - {\bf r}_j)\:,
\end{equation}
with 
\begin{eqnarray}
\label{force}
{\bf f} ({\bf r}) &\!\!\!\equiv\!\!\!&  - \sum_{\bf n} \frac{{\bf r} - {\bf n}L}{|{\bf
r}- {\bf n}L|^3} 
\Bigl[\,{\rm erfc}\left(\alpha | {\bf r} - {\bf n}L|
\right)  + \nonumber \\
&&  + 
\frac{2\alpha}{\sqrt{\pi}} |{\bf r} - {\bf n}L| \exp( -\alpha^2 | {\bf
r} - {\bf n}L|^2 ) \Bigr] - \nonumber \\
&& - \frac{1}{L^3}\sum_{\bf k} \frac{4\pi{\bf k}}{k^2} \exp
\left( -\frac{k^2}{4\alpha^2} \right) \sin\left( {\bf k}{\bf r}  \right) \:.
\end{eqnarray}
To achieve good accuracy with reasonable computational effort, typical
values are
$\alpha = 2/L$, $|{\bf r} - {\bf n}L| < 3.6 L$ and ${\bf k} \equiv 2 \pi
{\bf h} / L$ with ${\bf h}$ being an integer vector with $|{\bf h}|^2
< 10$ (see e.g. Hernquist et al. 1991; note, the second
component in their Eqn.~2.14b is missing a factor $ | {\bf r} -
{\bf n}L|$; it is identical to our Eqn.~\ref{force}).

For a particle pair with separation ${\bf
r}$, ${\bf f}({\bf r})$  includes the
contributions from all $L$-periodic	 	 
pairs with identical separation. To obtain the pure periodic
{\em correction force} ${\bf f}_{\rm cor}({\bf r})$, one has to subtract the direct
interaction of the central pair, which leads to
\begin{equation}
\label{corr-force}
{\bf f}_{\rm cor}({\bf r}) \:\equiv\: {\bf f}({\bf r}) + \frac{\bf
r}{\:|{\bf r}|^3}\:. 
\end{equation}
To obtain the periodic correction for the {\em potential} of the
particle pair, $\phi_{\rm cor}({\bf r})$, one proceeds similar and
again subtracts the isolated solution from the Ewald solution (Eqn.~\ref{potential}),
\begin{equation}
\label{corr-pot}
\phi_{\rm cor}({\bf r}) \:\equiv\: \phi({\bf r}) + \frac{1}{|{\bf r}|}\:.
\end{equation}
We compute the correction terms ${\bf f}_{\rm cor}$ and $\phi_{\rm
cor}$ for particle 
pairs on a Cartesian 
grid covering our whole simulation box, placing particle 1  in the
central node and particle 2 on different grid points, and obtain so a
table of pairwise force values.

\subsection{The PM method}
\label{pm-method}
Particle-mesh methods assign particle properties to mesh points,
solve the interaction equations on the grid and interpolate the solution
back onto the particles (see e.g. Hockney \& Eastwood 1988).
Typically, the density at each grid point is determined from the particle positions and
masses using CIC (``cloud in cell'') or TSC (``triangular shaped
cloud'') schemes, whose assignment functions are triangles or
quadratic splines, respectively (again Hockney \& Eastwood 1988, chap. 5). 
For the gravitational N-body problem one solves Poisson's equation.
Usually this is done in Fourier space, since the
differential operation there acts as a simple
multiplication. Therefore, the
density distribution on the grid is transformed into $k$-space using FFT and 
convolved with the appropriate Green's function to solve for the
potential. To obtain the forces, we convolve the Fourier transform of
the density with the Green's function for the force{\footnote{
Strictly speaking, a Green's function is the solution of an equation
of the type, $\nabla^2 {\cal G}({\rm r}) = -4 \pi \delta({\rm r})$ and
its boundary value problem. Green's functions are used to solve
Poisson's equation for the potential. Being the gradient of
Poisson's equation, the force equation can be solved by the same
ansatz.}}; separately
for the $x$, $y$ and $z$-component.
Finally, inverse FFT returns potential and force
at each grid point. The alternative 
for computing the forces in Fourier space is, to calculate them in real
space as gradients of the mesh-defined potential. This
would induce additional errors and therefore we prefer
the first method. 

\begin{figure*}
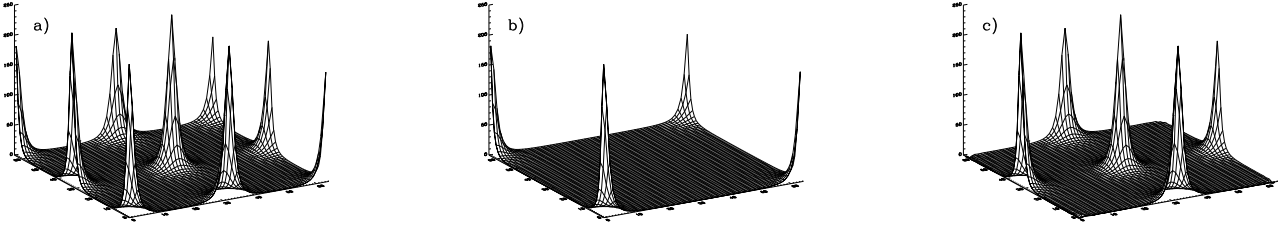

\centerline{
\epsfxsize=5cm \epsfbox{fig1a}\hfill
\epsfxsize=5cm \epsfbox{fig1b}\hfill
\epsfxsize=5cm \epsfbox{fig1c} }
\caption{Matrix of pairwise forces -- x-component $F_x$; cut through the
yz-plane at x=0: a) The periodic contribution (replicated onto the
whole grid). b) The isolated solution. c) The difference between both,
i.e. the correction term.} 
\label{fig-green}
\end{figure*}
\subsection{Combining the PM Method with Direct Summation on {GRAPE}}
In our application of the PM method, we compute the  (periodic)
{\em correction} to the force and the potential and add this to the
solution for the isolated system, 
which is obtained via direct summation on {\sc Grape}. This procedure ensures
proper treatment of periodic boundary conditions.
The Green's function for the PM scheme 
can be constructed directly in Fourier space (Hockney \& Eastwood
1988). However, for the correction force and potential, this  
is rather complicated and  it is more
intuitive and straightforward to proceed the following way:
We obtain the Green's function for each force component as the
Fourier transform of the $x$, $y$ and $z$-component, respectively,  of
the mesh-defined pairwise periodic correction force ${\bf f}_{\rm cor}$,
defined in Eqn.~\ref{corr-force}.  And for  the potential
correction, the appropriate Green's function is the Fourier transform
of $\phi_{\rm cor}$ in Eqn.~\ref{corr-pot}. The {\em corrective}
Green's function is thus 
the offset between the Green's function for the periodic system
$\cal{G}_{\rm per}$ and the isolated one $\cal{G}_{\rm iso}$,
\begin{equation}
\cal{G}_{\rm cor} \:=\: \cal{G}_{\rm per}- \cal{G}_{\rm iso}.
\end{equation}
Or with other words, ${\bf f}_{\rm cor}$ and
$\phi_{\rm cor}$ {\em are} the Green's functions with the right properties
and only have to be transformed into Fourier space for convolution
with the density.
Using the Fourier transform of the Ewald forces as periodic
Green's function was proposed by A.~Huss (private communication) and it
is straightforward to extend this for handling the force 
{\em correction} by subtracting the isolated solution.

Note, in order to obtain the {\em isolated} solution in Fourier space,
 one has to double the linear
dimensions of the grid and {\em zero pad} the additional grid points
to avoid contamination from implicitly assumed periodicity (see
Press et al. 1989). For example, to solve Poisson's equation for
a cubic density field with volume $L^3$, we have to use a grid of size
$(2L)^3$, assign the density field to one octant of the large box and
fill the  remaining 
grid points with zero. The Green's function, however, has to be defined on the
complete grid, i.e. ${\cal G}({\bf r}) = 1/|{\bf r}|$ for the
potential,  with $-L \leq x,y,z \leq L$. 
On the other hand, the {\em periodic} Green's function
is defined on the original grid, with $-L/2 \leq x,y,z \leq L/2$. For
alignment with the isolated solution, we have to extend the periodic
solution into the
larger cube. To obtain the right correction Green's function, e.g. for
the force, we 
replicate the table of pairwise Ewald forces
(defined on the small cube) into all octants of the large cube. Then
we subtract the isolated force field (defined on the large cube) and
transform into Fourier space. This procedure is illustrated in
Fig.~\ref{fig-green}, it plots $F_x$ in the yz-plane at $x=0$: a) is
the Ewald periodic force computed on a 32-grid and replicated four time into a
64-grid, b) is the isolated solution defined on the large grid and c)
is the difference, the final correction force. The Green's
function finally is the Fourier transform of this force
matrix. Convolution with the zero padded density field results in the
right force correction. The procedure is analogue for the potential.
In practice,  it is sufficient to compute the Green's function once at the
beginning of a simulation run and store it as a table.

The total force acting on an individual particle during one timestep
then stems from the particle system inside the simulation cube, computed
by direct summation on {\sc Grape}, {\em plus} the contribution from
an infinite set of periodically mirrored systems, computed via the
method described above. For smoothed particle hydrodynamics, one also has to add
pressure and viscous forces.  

We compute the periodic and the isolated solution on a grid, subtract the
latter from the first and add the isolated forces calculated with
{\sc Grape}. The two isolated solutions  cancel
out and one can ask the question, what have we gained? For a system with
more or less homogeneous density a pure PM scheme is
sufficient. However, the advantage of using {\sc Grape} is evident
when computing strongly structured systems.	
Unlike in PM schemes, the spacial resolution with {\sc Grape} is
not limited to a given cell size,  
but  adapts to the density distribution due to its Lagrangian nature.
It is limited only by the gravitational smoothing length, or equivalently,
by the choice of the minimum timestep. 
Since the potential of strong density peaks is dominated  by self
gravity and the influence of periodic boundaries (and thus of the
Ewald correction) becomes weak, it is sufficient to compute this correction
term on a relatively coarse grid which keeps the additional
computational cost low. For a relatively smooth density distribution,
a relatively wide mesh is sufficient anyhow.
The scheme proposed here unites both, high resolution with
{\sc Grape} and the periodicity of a PM scheme. 
In addition, applying a Courant-Friedrichs-Lewy like
criterion, we typically do not have to compute the FFT at each smallest
timestep, but can use stored correction values from the previous call.
This reduces the computational expense further.

\section{Optimization of the Scheme and Performance in Test Cases}
\label{opt-test}
The contributions to the total acceleration of one particle are
computed by two  distinct methods: by direct summation on the
{\sc Grape} board for the isolated solution and by applying a particle-mesh scheme
for the periodic correction term. Compared to the host computer, the
{\sc Grape} chips hereby have only limited
numerical accuracy (see Sect.~\ref{opt-resol}). The
spacial resolution of the PM scheme is limited by the number of grid zones and
the choice of the assignment function (Sect.~\ref{opt-assign}).
All this may lead
to spurious residuals when combining the forces;
some terms might not	 
cancel out completely. This ``misalignment'' can be
minimised  by randomly shifting the center of 
the simulation cube through the periodic particle distribution. Then
on average the force residuals cancel out and numerical stability is increased
(Sect.~\ref{opt-shift}). 
 
\subsection{Gravitational Smoothing Length and Numerical Accuracy of GRAPE}
\label{opt-resol}
\begin{figure}
\hspace{-0.3cm}\epsfxsize=8.5cm \epsfbox{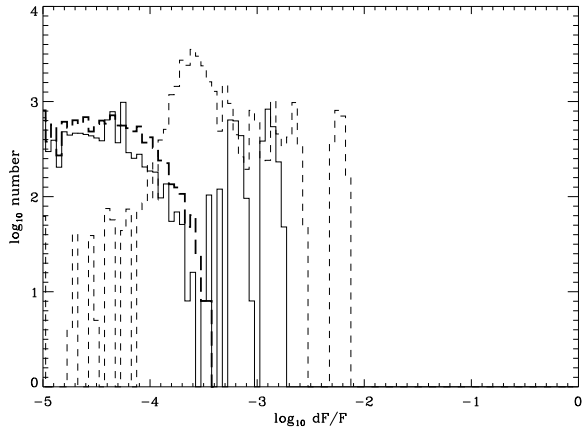}\hfill
\caption{Influence of the gravitational smoothing length
$\epsilon$. For all $\epsilon \le 0.001$ the force errors are equally 
distributed and minimal (thick dashed line). The errors grow with
increasing $\epsilon$ ($\epsilon = 0.005$ -- solid line, $\epsilon =
0.01$ dashed line).  
These values are computed for a distribution of $32^3$ particles on a
regular grid with periodic boundaries. The force correction is
computed on a $64^3$-mesh. The particles are centered on grid points
to be independent of the assignment scheme.}
\label{fig-epsilon}
\end{figure}

Since {\sc Grape} is used to calculate the force contribution from the
isolated system, we also call the board at the start of the
simulation, when setting up the force correction table. We subtract
the isolated solution from the periodic
Ewald solution for each particle pair on a grid.
For our test problems, we use a fixed gravitational smoothing
length $\epsilon$ throughout the 
whole computation and also for calculating the periodic Ewald forces. As
always in collisionless N-body simulations, the choice of $\epsilon$
is a trade-off between minimising 2-body relaxation  (large
$\epsilon$) and spacial resolution (small $\epsilon$). For the 
 simulations presented here, 
we typically adopt values between $\epsilon = 0.01$ and $0.001$, with
the total size of the simulation cube being $[-1,+1]^3$. 
The influence of $\epsilon$  is demonstrated in
Fig.~\ref{fig-epsilon}; it plots the force in x-direction, $F_x$, in
dimensionless units  for $32^3$ particles
which are distributed on a regular grid. Assuming periodic boundary 
conditions, this distribution is -- in principle -- force free.
For all values of $\epsilon \le 0.001$ the distribution of force errors is
equal and determined by the numerical accuracy  of the {\sc Grape}
chips. For larger $\epsilon$, the force contributions from the two
methods  get increasingly misaligned. 

\subsection{Grid Resolution and Assignment Function}
\label{opt-assign}
\begin{figure}
\hspace{-0.3cm}\epsfxsize=8.5cm \epsfbox{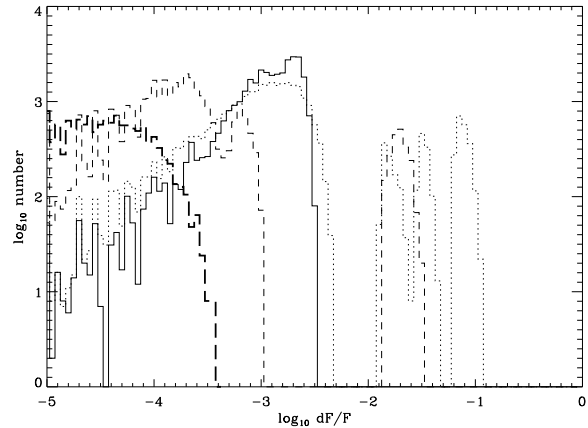}\hfill
\caption{Comparison of the relative force errors $dF/F$ for different
grid resolutions: $16^3$ (dotted line), $32^3$ (dashed line) and
$64^3$ (continuous line). As reference, the error distribution
stemming solely from {\sc Grape} is plotted as thick dashed line
(analogue to Fig.~\ref{fig-epsilon}).}
\label{fig-assign}
\end{figure}

Another important factor is the resolution of the grid. The more grid
zones are used, the smaller the wavelengths that  can be 
resolved by the PM scheme. However, the
number of CPU cycles increases linearly with the number of grid zones.
Again one has to find an compromise between accuracy and
computational speed, depending on the problem to be solved.
The errors furthermore depend on the adopted assignment function. To
obtain the 
force correction, one has to interpolate the particle distribution
onto a grid, solve Poisson's equation for the density field to obtain
forces  and assign these forces back onto the particles. One possibility is
to assign all particles to the cell they are located in, the nearest
grid point scheme, NGP. Another way is CIC (cloud in cell), which  uses a
boxlike cloud shape to distribute each particle into
eight neighbouring cells (in three dimensions).  TSC (triangular shaped 
cloud) distributes the particle mass into 27 cells 
 (for more details consult Hockney \& Eastwood 1988). For the problems
we study here, we consider NGP 
as too coarse, higher order schemes on the other hand smear out the
particle distribution  
and thus limit the spatial resolution. We adopt the CIC scheme in our
calculations. 
Using CIC, we plot in Fig.~\ref{fig-assign} the errors in $F_x$ 
for the distribution of $32^3$ particles on a regular periodic grid,
but now shifted by $dx = 0.015625$, i.e. half of the cell size of a
$64^3$-grid. As expected for the finer meshes, the interpolation is  more
accurate and the force errors are smaller. As a reference, we have
plotted the intrinsic error contribution coming from {\sc Grape} alone
(the thick long dashed line).
Note that the errors are not evenly distributed
amongst the particles; the errors are small and close to zero in the
interior of the simulated cube, whereas particles at the border have
errors by one or two orders of magnitudes larger (dotted
and dashed lines in the range $-2 < \log_{10} dF/F < -1$).
Note as well, that each grid is in fact  a factor of $2^3$ larger to
account for the zero padding necessary for the isolated solution,
i.e. the $64^3$-grid is in fact a $128^3$-grid. 

\subsection{The Influence of Random Shifts}
\label{opt-shift}
\begin{figure}
\hspace{-0.3cm}\epsfxsize=8.5cm \epsfbox{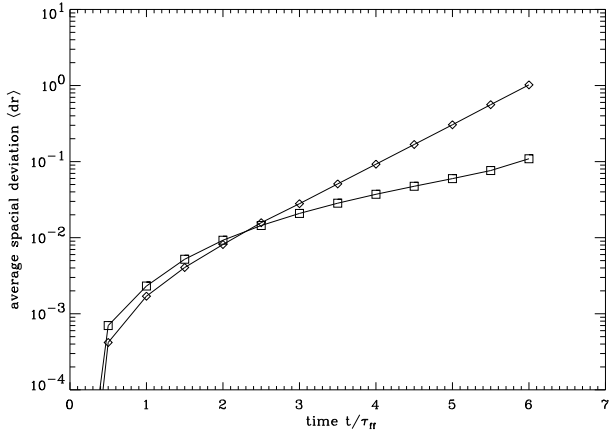}\hfill
\caption{Average particle displacement as function of time for the
system integrated with (squares) and without (diamonds) random shift.
}
\label{fig-shift}
\end{figure}

We can increase the stability of the code by minimising the influence
of the force errors for the border cells. 
One way to do this is 
to ensure, that these border cells do not always contain the same
group of particles: We apply a random shift of the whole simulation area. 
We can think of our simulation box as a window onto an infinite
periodic particle distribution. Since this distribution is infinite
and periodic we are free to choose any region in it, as long as it
contains the periodicity of the whole distribution. By randomly 
shifting the center of the simulation box we prevent the errors made
in the border cells to add up coherently for the same particles. On
average they cancel out. 
The stabilising effect of this method is demonstrated in
Fig.~\ref{fig-shift}. It compares the time evolution of the average
particle displacement (due to force errors) for the standard particle
distribution (particles on a regular $32^3$ mesh) for the two cases
with and without applying random shift. Clearly the system computed
without shift degrades much faster than the other one. In fact we plot
the pathological case, where the particles are located exactly between
two mesh points and therefore the assignment error is maximal. This
system collapses due to this errors on a timescale of about 5
free-fall times{\footnote{The regular infinite system does not
collapse; the free-fall time is per default infinite. Here and in 
the following free-fall time means the interval, an {\em isolated} simulation
cube would need to collapse.}}. For a system where the particles are placed exactly
on grid points the assignment errors are minimal. Such a system is
more stable without random shift. However, in a physically meaningful
context the particles are far from being placed exactly on grid
points and  random shift is an important tool of stabilising the code.

\subsection{Comparison with pure Ewald Method}
\begin{figure}
\centerline{\hspace{-0.3cm}\epsfxsize=8.5cm
\epsfbox{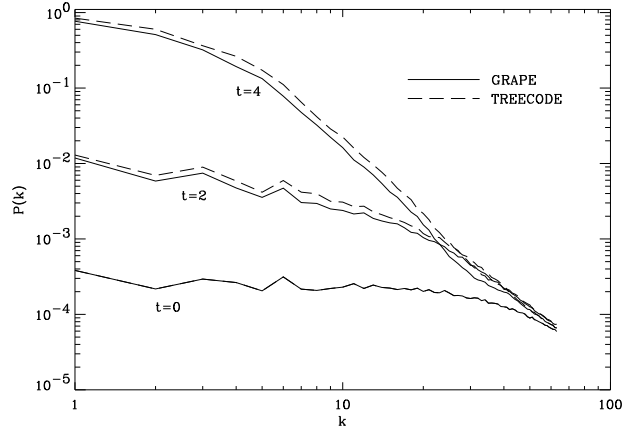}\hfill}
\centerline{
\hspace{-0.3cm}\epsfxsize=8.5cm \epsfbox{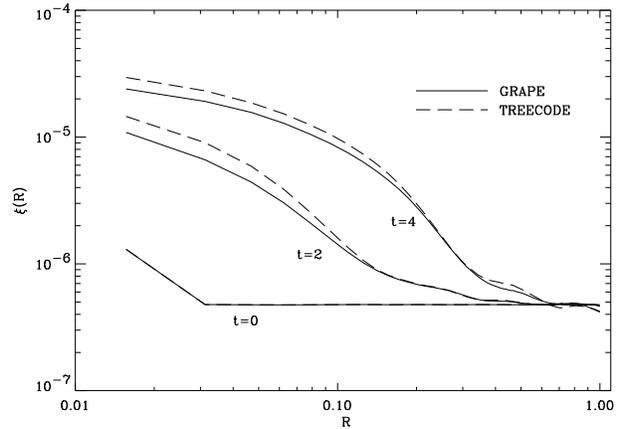}\hfill}
\caption{Power spectrum $P(k)$ and 2-point correlation function
$\xi(R)$ for the collapse of a random fluctuation field in periodic
boundaries. The solid lines describe the 
calculation done with {\sc Grape} and the dashed lines with a 
{\sc Treecode} scheme. Both systems evolve almost identically. Only for
small to intermediate particle separations, the difference in the
force calculation in both schemes leads to a deviation in the
correlation and in the power spectrum.}
\label{fig-power}
\end{figure}

A further test of the performance of this new method is, to compare it
with a {\sc Treecode} scheme. In this case, the Ewald method can be directly
implemented into the force summation, as described by Hernquist et
al. (1991). To do so, we distribute $32^3$ particles randomly
within the simulation cube to get a homogeneous distribution
with Poisson fluctuations. We chose the initial kinetic energy to be
zero and follow the fragmentation and collapse of the system. As
expected for this purely gravitational N-body system, density peaks
start to grow to form a network of filaments 
and knots, similar to those known from cosmological N-body
simulations -- note however, that we do not include a Hubble
expansion term. These knots finally merge to form one big clump.  
The evolution of the power spectrum $P(k)$ and of the correlation
function $\xi(R)$ is illustrated in
Fig.~\ref{fig-power}. 
Considering the differences in the force calculation method, the
evolution of the two systems agree remarkably well. 

\section{Fragmentation of Molecular Clouds}
\label{mol-clouds}
As a first application of the method presented above we describe the
simulation of the dynamical evolution and fragmentation processes  in
the interior of giant molecular clouds. 
\begin{figure*}
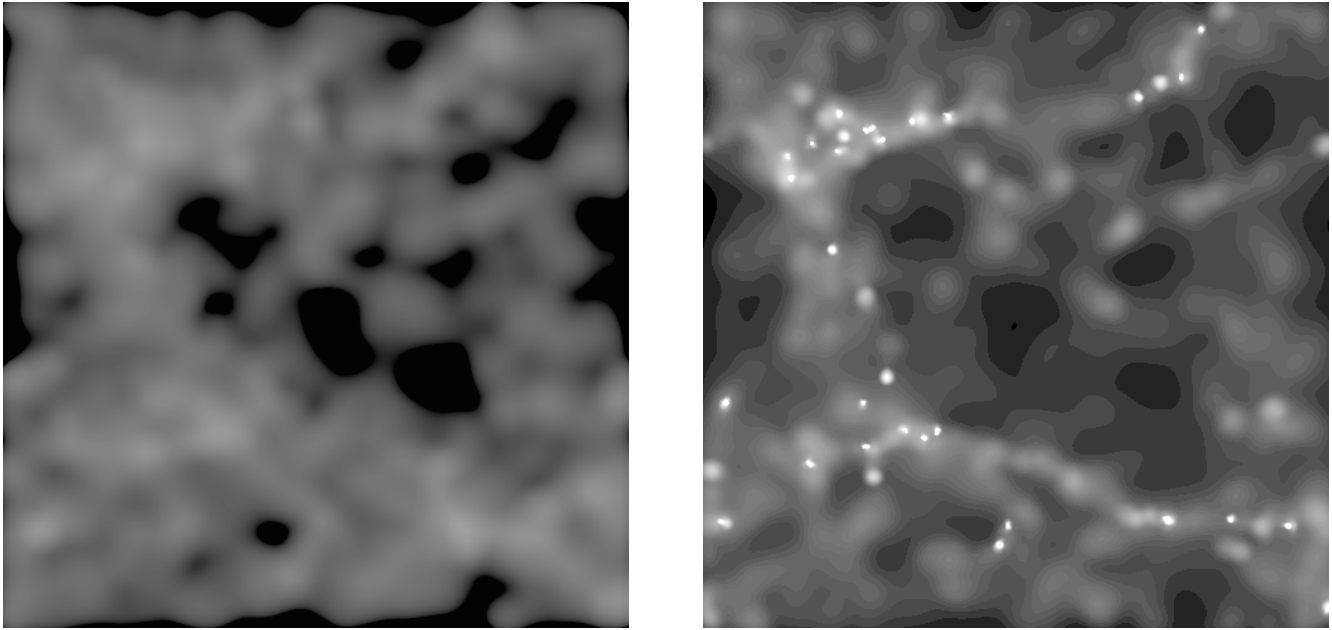

\centerline{
\epsfxsize=8.3cm \epsfbox{fig6a} \hfill
\epsfxsize=8.3cm \epsfbox{fig6b}}
\caption{Evolution of a Gaussian density field with $P(k) \propto
1/k^2$: Initial condition (left side) and after one free-fall time (to the right). The
density is normalised equally and scaled logarithmically in both
images. The simulation was done with fully periodic {\sc GrapeSPH}
code using $100\,000$ particles.} 
\label{fig-simul}
\end{figure*}

Molecular clouds are highly structured:
Observations reveal a hierarchy of filaments and sheets, clumps and
sub-clumps, ranging from scales of the size of the whole 
cloud (see e.g. Bally 1996) down to the smallest objects resolved
by todays telescopes (Wiseman \& Ho 1996). 
Density-size or linewidth-size relations indicate that typically a
cloud is to a large extent supported by 
``turbulent motion''; the measured linewidths exceed the thermal
line broadening by far. Observations of Zeeman splitting and
polarisation indicate the presence
of magnetic fields. Furthermore feedback mechanisms from 
newly formed stars, outflows, stellar winds, ionization fronts and finally
supernovae, produce shells and bubbles and deposit huge amounts
of energy and momentum into the interstellar medium. The physical
processes in molecular clouds are extremely diverse. A
comprehensive overview can be found in ``Protostars and Planets III'',
eds. Levy \& Lunine, 1993)

However, this is the environment in which stars form. To assess the
problem of fragmentation and clump formation that leads to star
formation, we start in the most basic way: We follow the dynamical
evolution of isothermal gas in a region in the interior of a molecular
cloud, starting from an initial density distribution to the formation
of selfgravitating clumps. We solve the hydrodynamical equations using
the SPH with {\sc Grape} and  implement periodic boundary conditions
as described above to prevent global collapse. 
Ignoring additional physical effects, we 
study the interplay between gravity and gas pressure, 
which  by itself will 
produce hierarchical filamentary structure, as was derived by de Vega,
Sanchez \& Combes (1996).  Figure \ref{fig-simul} describes one simulation:
The left image is the projection of the initial
density field, that was assumed to be Gaussian with a $1/k^2$ power 
spectrum. The right image depicts the same field after one free-fall
time{\footnote{Again, free-fall time means the time interval an
{\em isolated} system would need to collapse.}.
The initial Gaussian fluctuations evolved into a system of
filaments and knots, some of which contain collapsing cores.
The initial conditions in Fig.~\ref{fig-simul} where chosen to be
very cold; the total mass in the system exceeds the Jeans
mass by far and a large number of initial density peaks are
selfgravitating and begin to collapse. Gravity dominates strongly over
pressure forces. 
%The situation of a very cold system is thus quite similar to
%cosmological simulations of galaxy formation (besides the fact that
%it neglects the Hubble expansion).

Figure~\ref{fig-mass} shows the mass distribution of the identified
clumps. We divide the mass range into equal logarithmic intervals and 
 plot the number of clumps identified, divided by the
interval. For masses below $m \simeq  0.001$, our clump finding
algorithm deteriorates and the sampling gets incomplete.
 With the chosen initial conditions and parameters, the mass
distribution follows a power law, $dN/dm \propto m^n$, with $n
\simeq -1.5$, which is indicated by the dashed line. This is in 
agreement with the values observed in molecular clouds (see e.g. Blitz
1993) although the observations exhibit a huge scatter.
These results of our study of the fragmentation process in 
molecular clouds are still  preliminary; more details  will be
discussed in a subsequent paper (Klessen \& Burkert 1997).

Certainly this treatment is very coarse. It neglects the
presence of magnetic fields and the input of energy and momentum by
young stars. Future numerical studies of the evolution of interstellar medium and the
formation of stars have to take these processes into account.
However, the simple isothermal model presented here is able to 
reproduce some of the observed properties of giant molecular clouds
and star forming regions. 
\begin{figure}
\epsfxsize=8.5cm \epsfbox{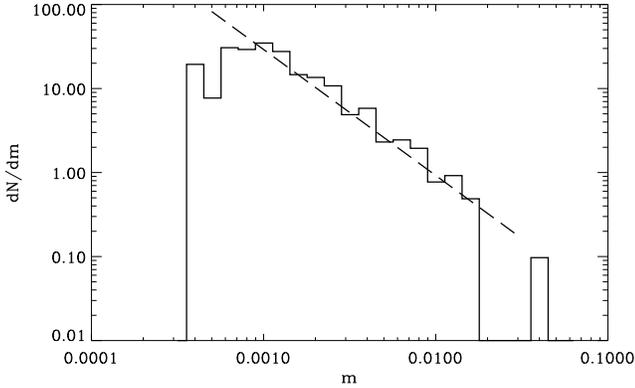}\hfill
\vspace{-0.3cm}
\caption{Mass distribution of the identified clumps. The dashed line
indicates the slope $dN/dm \propto 
m^{-1.5}$. The total mass in the system is $m=1$; for clump masses
$m<0.001$ the sample gets incomplete.}
\label{fig-mass}
\end{figure}

\section{Summary} 
\label{summary}
We have presented a method of how to incorporate periodic boundary
conditions into numerical simulations using the special hardware
device {\sc Grape}. We compute a periodic correction force onto each
particle applying a particle-mesh like scheme: The particle
forces in the isolated system are calculated via direct summation on the {\sc Grape}
board. Then we assign the particle distribution to a mesh and compute
the periodic correction force for each grid point by convolution with
the appropriate Green's function in Fourier space. These forces are
assigned back onto the individual particles. This method unites the
advantages of {\sc Grape} as being Lagrangian and offering good
spacial resolution necessary for treating highly structured density
distributions with the advantage of particle-mesh schemes as being
intrinsically 
periodic. The potential of strong density peaks is
dominated  by the local mass accumulation. In such a region the
influence of the global properties of the system is
week and it is therefore sufficient to compute the correction 
term on relatively coarse grid. This keeps the additional
computational cost low. 

The method proposed here has proved to be numerically stable
and inexpensive in terms of computer time. Thus the
special purpose hardware device {\sc Grape} can be applied
successfully to astrophysical problems that require the correct treatment
of periodic boundary conditions. As an example we study the
dynamical evolution and fragmentation process in the interior of
molecular clouds.

\section*{Acknowledgments}
I am grateful to Andreas Huss for many
stimulating discussions. I thank Ralph-Peter Anderson, Matthew Bate and
Matthias Steinmetz for many useful suggestions and Andreas Burkert
for his critical remarks and advice, and for carefully reading this manuscript.

%%%%%%%%%%%%%%%%%%%%%%%%%%%%%%%%%%%%%%%%%%%%%%%%%%%%%%%%%%%%%%%%%%%%%%
\end{document}